\documentstyle[art8,aas2pp4,flushrt,tighten]{article}

\slugcomment{Submitted to Astrophysical Journal}

\lefthead{Shlosman \& Heller}
\righthead{Nested Bars in Disk Galaxies}

\begin{document}

\title{NESTED BARS IN DISK GALAXIES:\\ NO OFFSET DUST LANES
IN SECONDARY NUCLEAR BARS}

\author{Isaac Shlosman\altaffilmark{1,2}}

\affil{Joint Institute for Laboratory Astrophysics, University of Colorado,
    Box 440, Boulder, CO 80309-440 \\ email: {\tt shlosman@pa.uky.edu}}

\and

\author{Clayton H. Heller}

\affil{Department of Physics, Georgia Southern University, Statesboro, GA
      30460-8031 \\ email: {\tt cheller@gasou.edu}} 

\altaffiltext{1}{JILA Visiting Fellow}
\altaffiltext{2}{permanent address: Department of Physics and Astronomy, 
University of Kentucky, Lexington, KY 40506-0055}

\begin{abstract}
Under certain conditions, sub-kpc nuclear bars form inside large-scale
stellar bars of disk galaxies. These secondary bars spend a fraction of their
lifetime in a {\it dynamically-decoupled} state, tumbling in the gravitational
field of the outer bars. We analyze the flow pattern in such nested bar
systems under the conditions of negligible self-gravity and find that secondary
bars differ fundamentally from their large-scale counterparts, in gas flow
pattern and other dynamical properties. In particular the gas flow across the
bar-bar interface in these systems can be more chaotic or more regular in 
nature, and, contrary to predictions, has no difficulty in penetrating the
secondary bars along the primary large-scale shocks. The outer parts of both
short and long  nuclear bars (with respect to their corotation) appear to be
depopulated of gas, while deep inside them the flow exhibits low Mach numbers 
and follows ovally-shaped orbits with little dissipation. Long nuclear bars
remain gas-rich longer, and for this, relatively short, period of time are
largely of  a rectangular shape, again with a small dissipation. We find that
gas-dominated and star-dominated nuclear bars 
avoid the bar-bar interface, making both types of bars short relative to their
corotation.
Furthermore, our  earlier work has shown that {\it dynamically-coupled}
secondary bars exhibit a similarly relaxed low-dissipation flow as well.
Therefore, no large-scale shocks form in the nuclear bars, and consequently,
no offset dust lanes are expected there. We find that offset dust lanes cannot
be used in the search for secondary (nuclear) bars. Finally, we discuss the
importance of gas self-gravity in the further evolution of these systems.
\end{abstract}


\section{Introduction}

Double bars appear to be frequent among disk galaxies, probably in excess of
20-25\% (e.g., Friedli et al. 1996; Erwin \& Sparke 1999; Laine et al. 2001).
About 1/3 of barred galaxies host a second bar (Laine et al.). They
consist of large-scale stellar (primary) bars and sub-kpc (secondary) nuclear
bars. Examples of such systems have been known since de Vaucouleurs (1974),
Sandage \& Brucato (1979) and Kormendy (1983), although they had attributed
the twisting of the innermost isophotes to a triaxial bulge. Dynamical
consequences of nested bars for the stellar and gas dynamics in disk galaxies
have been studied both theoretically and numerically, but are far from being
understood (Shlosman, Frank \& Begelman 1989; Pfenniger \& Norman 1990;
Friedli \& Martinet 1993; Combes 1994; Heller \& Shlosman 1994; Maciejewski
\& Sparke 2000). They have been also implicated in the fueling of active
galactic nuclei and starburst activity in the central kpc (Shlosman, Begelman
\& Frank 1990; Athanassoula 1994; Friedli 1999).

One of the most interesting aspects of nested bars is their dynamically
decoupled phase, when the rate of tumbling of each bar is different. Shlosman
et al. (1989) have shown that self-gravitating gas instabilities within the
central kpc can be the prime reason for this runaway process, which was
confirmed in numerical simulations (Friedli \& Martinet 1993, Combes 1994;
Heller \& Shlosman 1994; Shlosman 2001).
Furthermore, Shaw et al. (1993) and Knapen et al. (1995b; see also Shlosman 1996)
have studied gas accumulation in the vicinity of the inner Lindblad
resonance of large-scale primary bars which manisfested itself in the formation
of gaseous and stellar secondary bars still {\it coupled} to the background
potential and tumbling with the primary bar pattern speed. Pfenniger \& Norman (1990) 
used weakly dissipative equations of motion for a test particle in a double bar
potential. Heller, Shlosman \& Englmaier (2001) found that the formation and
decoupling of the secondary gaseous bar is possible even in the limit of
weak self-gravity in the gas. Finally Maciejewski \& Sparke (2000) have
invoked multi-periodic (loop) orbits to support the time-dependent gravitational
potential of a double bar system.

Detection techniques of secondary bars depend on the type of the bar, namely
gaseous or stellar. The stellar bars have been found using photometry,
most efficiently in the optical and NIR (e.g., Scoville et al. 1988; Buta \&
Crocker 1993; Knapen et al. 1995a; Shaw et al. 1995; Friedli et al 1996; 
Jungwiert, Combes \& Axon 1997; Erwin \& Sparke 1999; Knapen, Shlosman \&
Peletier 2000; Laine et al. 2001). The gaseous molecular bars have been
observed using CO and H$_2$ (e.g., Ishizuki et al. 1990; Devereux, Kenney \&
Young 1992; Maiolino et al. 2000; Sakamoto, Baker \& Scoville 2000). Recently
Regan \& Mulchaey (1999)  and Martini \& Pogge (1999) have invoked the 
so-called offset dust lanes to
compare the frequency of secondary bars in Seyferts and `normal' galaxies,
using high-resolution HST imaging. Such offset lanes are routinely detected in
large-scale bars and represent global (collisional) shocks in the ISM, in
response to torquing by the bar potential. Extensive numerical study has
revealed the close connection between the shape of these dust lanes, mass
distribution in the galaxy, strength of the stellar bar and its pattern speed
(Athanassoula 1992).  

In this paper we analyze the gas dynamics in dynamically-decoupled secondary 
nuclear bars and show that the gas response to their torquing is
fundamentally different from that of their large-scale counterparts. This
leads to a number of  theoretical and observational consequences. Our results
clarify the details of gas flow across the bar-bar interface and within the
secondary bar. We find that no large-scale shocks and offset dust lanes can
form in the nuclear bars. Dust lanes, therefore cannot be used to search for
nested bar systems, in general, and, specifically, cannot address the issue of
morphological differences between Seyfert and normal host galaxies.

\section{Model Construction and Orbit Analysis}

\subsection{Orbits in single and nested bars}

To study the gas flow in the nested bar systems, we constructed a grid of
models, with two representative cases analyzed below. We use the 2-D version
of our Smooth Particle Hydrodynamics (SPH) code (Heller \& Shlosman 1994) in
the background potentials of both bars embedded in the disk and spheroidal
components, neglecting the gas self-gravity. This code has a dynamic spatial
resolution which is defined by the kernel smoothing length. An SPH neighborhood 
of 96 gas particles was used. 

In this section we define the 
model potentials, examine the orbits they support and verify the positions of
inner resonances in the disk by means of nonlinear orbit analysis (see Heller
\& Shlosman 1996 for technical details).  This is important because these
resonances not only describe the distribution of main families of periodic
stellar orbits but provide an insight into the gas response to nested bar
torquing. 

Two resonances, corotation and the inner Lindblad resonance(s)\footnote{The
ILRs are resonances between the stellar orbital precession frequency and the
bar pattern speed.} (ILRs), play a special role in the evolution
of disk galaxies with {\it single} bars. Two families of periodic orbits are
dominant within the bar corotation. The first family, so-called $x_1$, is
aligned with the bar and extends between the center and the corotation radius
(Contopoulos \& Papayannopoulos 1980). The second family, $x_2$, is found
between the ILRs and its orbits are oriented perpendicular to the bars major
axis. These latter orbits weaken the bar when they are populated.

Both $x_1$ and $x_2$ families are resonant orbits which tumble with the
bar pattern speed. Each of these orbits corresponds to a fixed Jacobi energy,
$E_J$, which is a constant of motion along the orbit in the rotating frame
of the bar (e.g., Binney \& Tremaine 1987). The orbital shapes do change with
$E_J$. 

Gas orbits of course do not follow exactly either of the two main families of
orbits because of internal dissipation and do not conserve $E_J$. This energy
non-conservation is amplified further for orbits with loops or  pointed
ends, in which case shocks develop and the gas rapidly depopulates them. In
the bar frame of reference the gravitational potential of a
single-barred galaxy is time independent, which makes it easier to describe
the gaseous response. Unlike the stellar orbits which can change their
response abruptly at each resonance, from being aligned with the bar to being
perpendicular, and vice versa, the gas responds gradually and its orbits
change their orientation by forming a pair of offset shocks. These shocks have
been detected by the dust lanes, whose shape is constrained by
the ratio of corotation-to-bar-radii, $1.2\pm 0.2$, as found empirically by
Athanassoula (1992). 

In the {\it nested} bar galactic systems,  when both bars are dynamically
decoupled and tumble with different pattern speeds, the gravitational
potential is time-dependent in all frames of reference. In such a case Jacobi
energy is not an integral of motion even for the collisionless `fluid,' i.e.,
stars. One can look for an alternative treatment such as the one proposed by
Maciejewski \& Sparke (2000) who introduced multiperiodic orbits, called
`loops,' supporting the double bar system. However, it is not clear what
fraction of the phase space is occupied by these loops and how many orbits are
actually trapped around them. In any case these orbits are not
suitable for the gas, as all of them are intersecting. 

In order to understand the gas flow in a nested bar potential, we analyze
the main families of periodic orbits in the frames of reference of each of the
bars, with the other bar being symmetrized. This allows one to interpret the
observed gas response in the numerical simulations of these systems, albeit 
roughly. 

A number of rules need to be followed in order to construct a self-consistent
nested bar system. First, a necessary condition is to accumulate a critical
mass of gas, which initiates the runaway. Although in this work we assume a
system already in a decoupled state, it still requires the existence of an ILR
in the primary bar for consistency. This constrains the pattern speed of the
primary bar. Second, the corotation radius of the secondary bar must be found
in the vicinity of the above ILR, in order to decrease the fraction of chaotic
orbits generated at each resonance. This fixes the pattern speed of the
secondary bar. Third, the length of a secondary bar cannot exceed its
corotation, but unlike in the primary bars, it can be substantially smaller.
This length is determined by the amount of dissipation in the gas at the time
of gas settling inside the ILR on the $x_2$ orbits and its subsequent
dynamical runaway. The details of this process are outside the scope of this
paper (see e.g., Heller et al. 2001; Shlosman 2001; and in preparation).

\subsection{Building the model}

Following the above constraints, we have chosen two models which differ
mainly due to the length of the secondary bar. Model~1 has a relatively short
secondary bar, confined well inside its corotation (and the ILR of the primary
bar). Model~2 hosts a secondary bar extending to its Ultra-harmonic resonance,
which is located at about 0.83 of its corotation radius and is typical of the
primary or single bars, as discussed in Section~1.

The disk and bulge/halo potential which is identical for both models is given
by a Miyamoto \& Nagai (1975) analytical model,

$$  \Phi = - {GM\over \sqrt{r^2+(A+B)^2}},  \eqno(1)  $$
where $M$ --- mass in units of $10^{11}~{\rm M_\odot}$ and $A+B$ ---  scaling
parameters representing the disk radial scalelength, in units of 10~kpc
(Table~1). Each bar is represented by Ferrers (1877) potential with $n=1$,
with the primary bars being identical in both models. In dimensionless units,
nested bar masses and semi-major ($a$) and semi-minor ($b=c$) axes are given in
Table~2.  The axial ratio for the primary and secondary bars is $b/a \sim
1:3$ and 1:4, respectively, which corresponds to moderately strong to
strong bars. The dynamical time at 10~kpc, $\tau_{\rm dyn}=1$, corresponds to
$4.7\times 10^7$~yrs.  

\begin{table}
\caption{Model Potential Parameters}
\smallskip
\begin{tabular}{lccc} \hline\hline
Component & Mass & A & B \\ \hline
disk      & 0.6  & 0.2  & 0.05  \\
bulge     & 0.03 & 0.01 & 0.02  \\
halo      & 1.0  & 0.0  & 1.0   \\ \hline
\end{tabular}
\end{table}

\bigskip\bigskip
\begin{table}
\caption{Ferrers Bar Parameters ($n=1$)}
\smallskip
\begin{tabular}{lccccc} \hline\hline
Model & Bar & Mass & $a$ & $b$ & $\Omega$ (bar)\\ \hline
1   & primary   & 0.15   & 0.65  & 0.22   & 1.0 \\
    & secondary & 0.007  & 0.05  & 0.0125 & 8.3 \\ \hline
2   & primary   & 0.15   & 0.65  & 0.22   & 1.0 \\
    & secondary & 0.012  & 0.075 & 0.0187  & 8.3 \\ \hline
\end{tabular}
\end{table}

In both models presented here the primary bar ends at $r\approx 0.65$, just
inside its Ultra-Harmonic resonance at 0.7. The corotation radius of the
secondary bar, as discussed above, lies near the outer ILR of the primary bar, 0.09
and 0.10, for Models~1 and 2. Finally, each bar comprises about 20\% of the
total mass {\it within their respective radii}. Initially, the gas is
distributed exponentially, with a scalelength of 0.3, and its sound speed is
$15~{\rm  km\ s^{-1}}$.     

For comparison, we also present Model~3, whose radial mass distribution
is identical to that of Model~1, but with an axisymmetrized secondary bar.
Dynamical effects of secondary bars are emphasized this way.        

\figurenum{1}
\begin{figure}[ht!!!!!!!!!!!!!!!!!!!!!!!!!]
\vbox to4.4in{\rule{0pt}{4.4in}}
\includegraphics{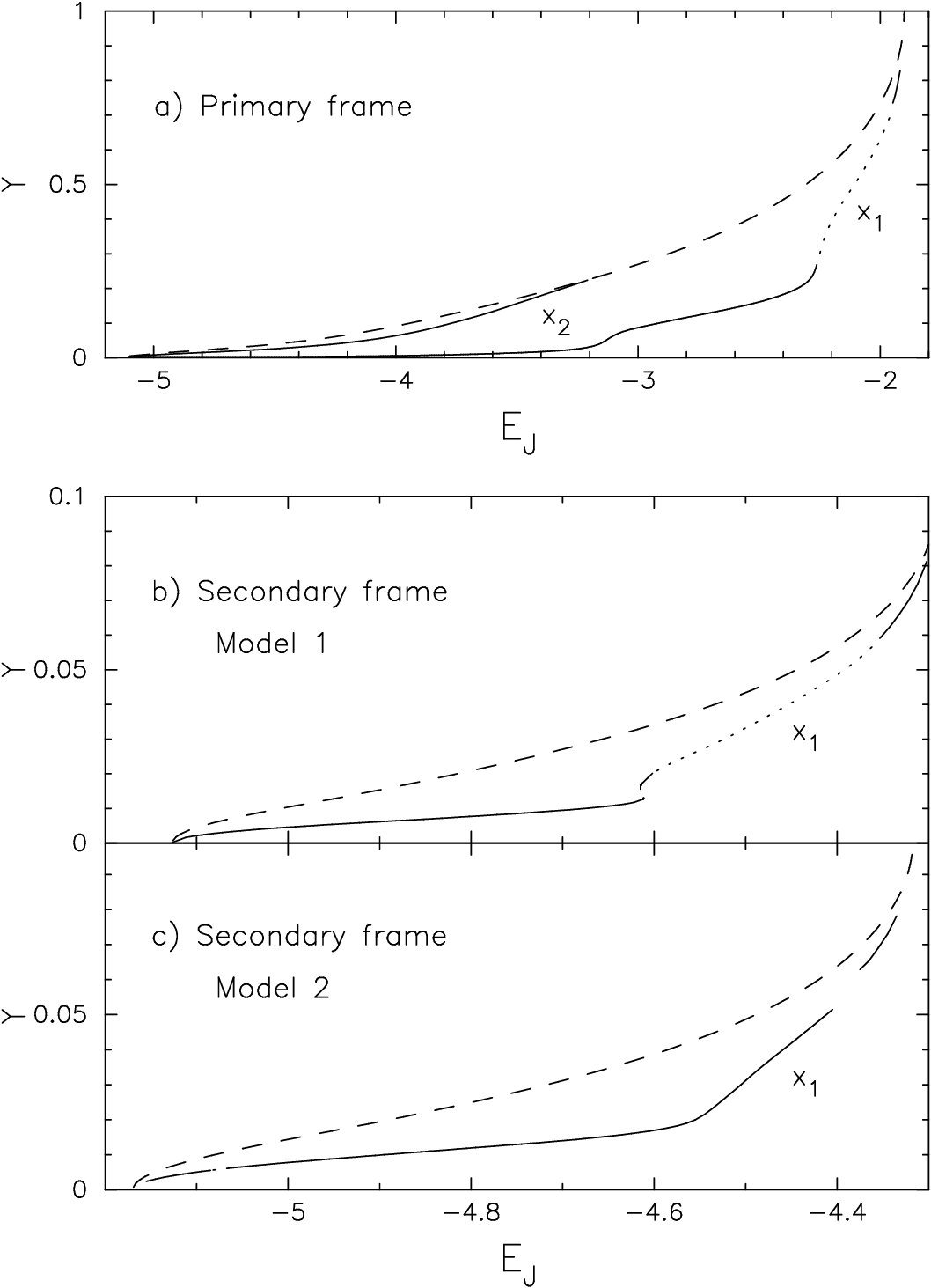}
\caption{Characteristic $x_1$ and $x_2$ orbit diagram for Model~1 (top, middle) and
Model~2 (lower). The upper panel ($a$) gives the orbit analysis in the frame
of the primary bar, with the secondary bar being axisymmetrized. The middle
($b$) and lower ($c$) panels give the analysis in the frame of the secondary
bar with the primary bar  axisymmetrized. The dashed line is   the zero
velocity curve. The dotted line represents unstable $x_1$ orbits. The galactic
center is on the left.}
\end{figure}           

\subsection{Nonlinear orbit analysis}

The choice of the primary bar pattern speed, $\Omega_p=1.0$, and mass
distribution has led to a double ILR in all models. The existence and positions
of these resonances was verified using nonlinear orbit analysis.  Fig.~1a
which is the characteristic diagram for both Models~1 and 2 in the frame of
reference of the primary bar, $\Omega_p$, allows one to understand  the degree
of orbital support for this bar. The $x_1$ orbits extend from the center till
the corotation, while the $x_2$ orbits are limited between the ILRs. For a
fully developed secondary bar, the inner ILR is basically located at the very
center of the disk, at $r\sim 0.008$ along the x-axis (0.01 along the y-axis),
i.e., its dynamical effect is not important here. Instead, we expect the
perturbation of the secondary bar to dominate the dynamics at these radii.
At the same time the outer ILR of the primary bar is found along the primary
bar major (minor) axis at $\approx 0.08$ (0.10) (Model~1), and at $\approx
0.09$ (0.24) (Model~2). To estimate the location of the primary ILR, the
secondary  bar was axisymmetrized.               

The characteristic diagrams presented in Fig.~1, especially those in the
frame of reference of the secondary bar, hint about the gas response observed
in our numerical simulations. No $x_2$ orbits exist in the frame of  reference
of the secondary bar which tumbles fast enough to avoid its ILRs 
(Figs.~1b,c). In Fig.~1b (Model~1), a shoulder in the $x_1$ characteristic  is
visible. Note that a broad range of these orbits at higher Jacobi energies is
unstable, the exact reason for  which will be discussed elsewhere. Moreover,
these unstable orbits intersect with the $x_1$ orbits at lower energies (deeper
inside the bar), which corresponds to spatial scales outside $r\sim 0.03$.
This itself means that the gas will not be able to settle down in the outer
half of the secondary bar in Model~1.

The  corresponding characteristic diagram produced for Model~2 (Fig.~1c) is
very different. First, no shoulder exists in the $x_1$ characteristic,
which are non-intersecting and stable all the way till the corotation of this
bar. One would expect the gas to fill up completely the secondary bar under
these conditions. This indeed happens (Section~3) before the gas is driven further
inwards due to the time-dependent potential. The overall difference between
Models~1  and 2 is due to the larger quadrupole moment of longer secondary bar in
the latter Model, which extends up to  the Ultra-harmonic resonance, near its
corotation.

\section{Model Evolution: Gas Flow in Nested Bars}

To avoid transients, the nonaxisymmetric potential is turned on gradually. The
resulting gas response to the nested bar potential in both models can be
best inferred by defining three regions in the disk. Namely, ($i$) that of the
primary bar (outside its ILR region), ($ii$) the bar-bar interface (hereafter the
{\it interface}) encompassing the outer ILR of the large bar and the
outer part of the secondary bar, and ($iii$) the interior of the secondary bar. 

\figurenum{2}
\begin{figure}[ht!!!!!!]
\vbox to4.7in{\rule{0pt}{4.7in}}
\includegraphics{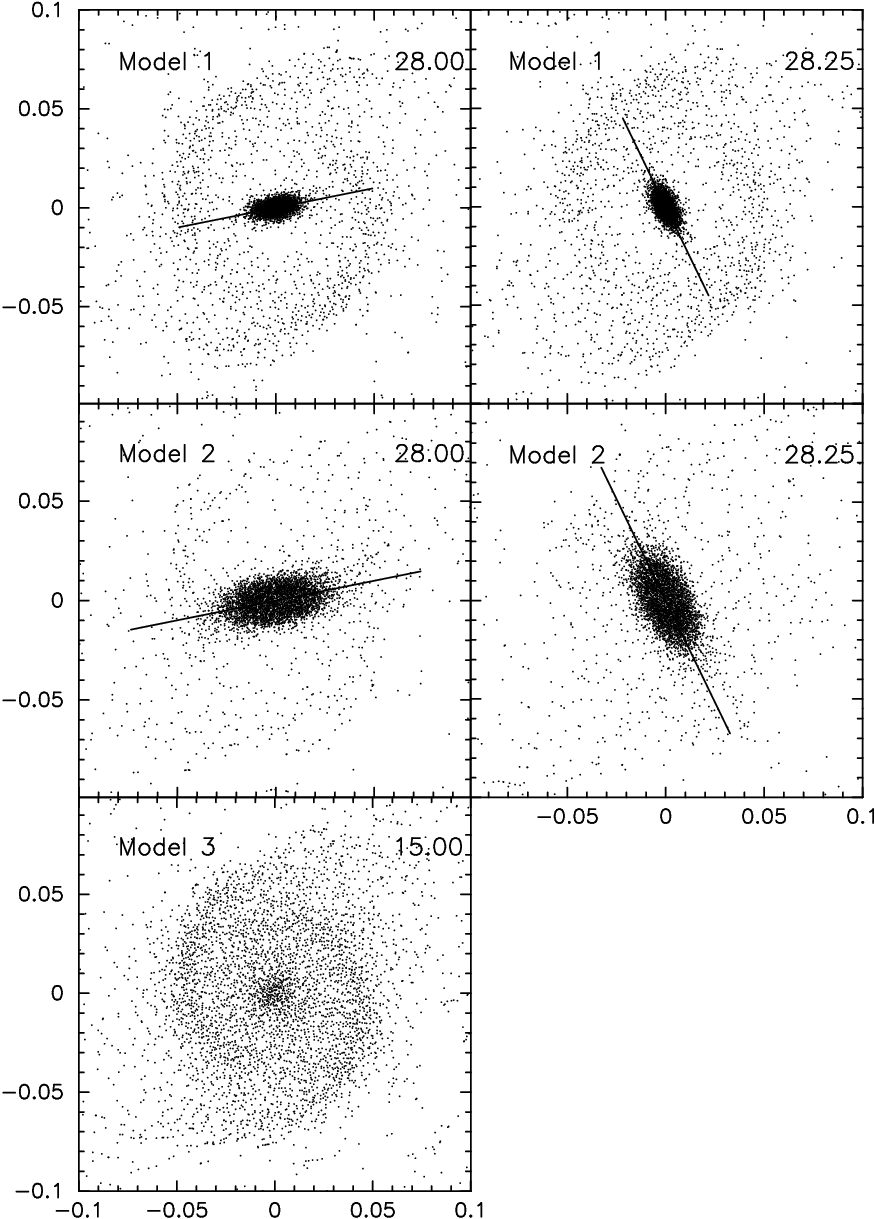}
\caption{Time evolution of the gas density distribution in the central
kpc ($r=0.1$) in Models~1--3: 2D SPH simulation in the background gravitational
potential of a nested bar disk galaxy (face-on). The gas response to the
nested bar torquing is shown in the primary bar (horizontal) frame of
reference. Both bars and gas rotate counter-clockwise. The position angle of
the secondary bar and its length are indicated by a straight line. Time is given
in units of dynamical time $\tau_{dyn}$. Note the absence of a nuclear ring
in Model~2.}
\end{figure}      
\figurenum{3a}
\begin{figure}[ht!!!!!!]
\vbox to2.8in{\rule{0pt}{2.8in}}
\includegraphics{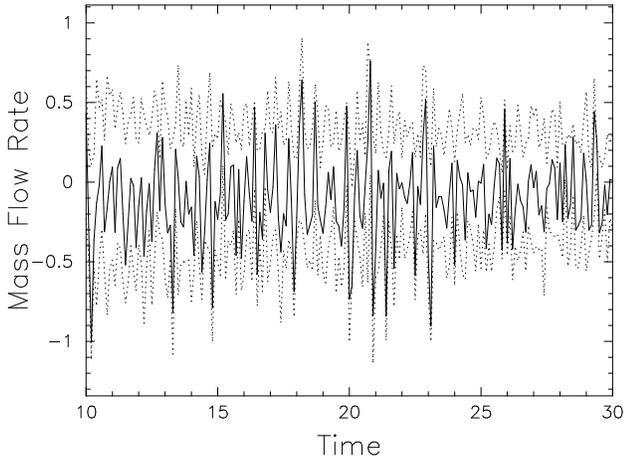}    
\caption{Time evolution of the gas inflow (negative) rate across the bar-bar
interface, $r=0.1$ (i.e., corotation of the secondary bar), of a double bar
system (Model~2). {\it Upper dotted line:} flow rate within $\pm 45^\circ$ of
the major axis of the primary bar. {\it Lower dotted line:} flow within $\pm
45^\circ$ of the minor axis of the primary bar. {\it Solid line:} total
flow across the bar-bar interface. Time is given in units of dynamical time
$\tau_{dyn}$.}
\end{figure}         
\figurenum{3b}
\begin{figure}[ht!!!!!!]
\vbox to4.5in{\rule{0pt}{4.5in}}
\includegraphics{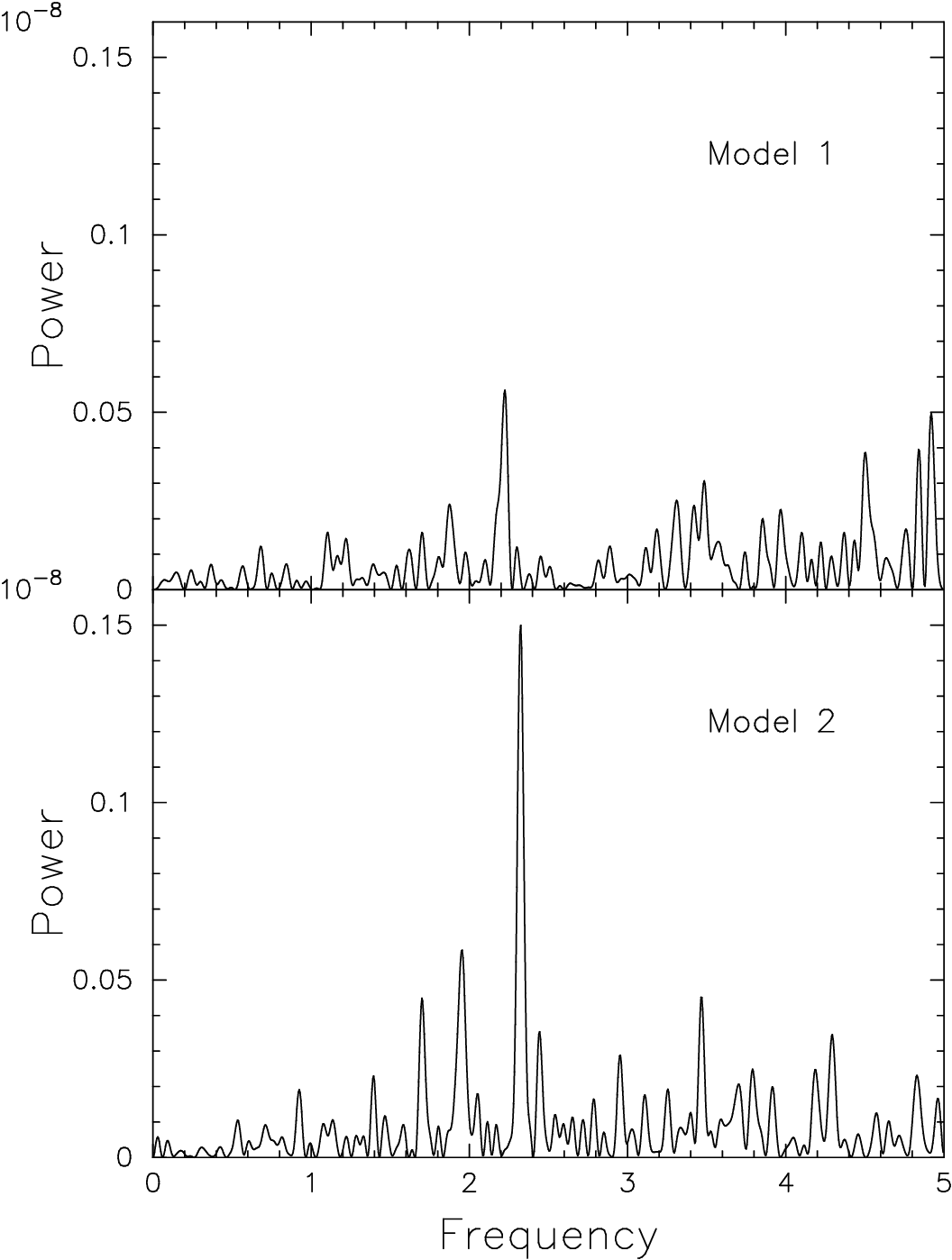}   
\caption{The Fourier transform of the total gas inflow rate across the bar-bar 
interface, $r=0.1$. Significant power at the beat frequency of the two bars,
$f\sim 2.3$, can be seen for Model~2. Model~1 also exhibits this beat but with
substantially less power, along with additional power at higher frequencies.}
\end{figure}   
\figurenum{4a}
\begin{figure}[ht!!!!!!!!!!!!!!!!!!!!!!!!!]
\vbox to5.7in{\rule{0pt}{5.7in}}
\includegraphics{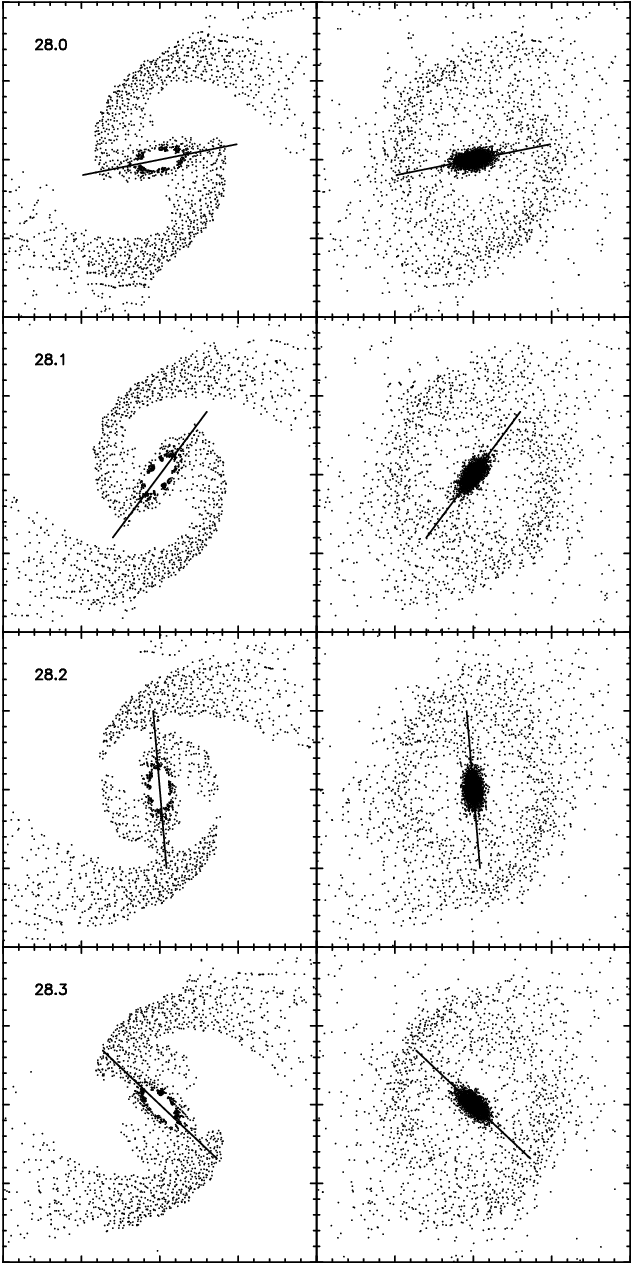} \caption{Pattern of shock dissipation (left) and density evolution
(right) in the central kpc ($r=0.1$) of Model~1, shown in the frame of
reference of the primary bar (horizontal). Positions of the secondary bar and
its length are indicated by a straight line. All rotation is
counter-clockwise. The particles on the left are those having greater than
average dissipation rate which is given by the time derivative of the
nonadiabatic component of internal energy. Note the sharply reduced
dissipation in the innermost 0.02 and ``limb brightening'' enveloping it. Also
visible are two dissipative systems associate with the large-scale shocks in
the primary bar and with the trailing shocks in the secondary bar.}
\end{figure}

In the first region, the primary bar outside the interface, the gas responds
by forming a pair of large-scale shocks, corresponding to the offset dust lanes,
and flowing inwards across the interface into the secondary bar, for Models~1
and 2, while stagnating in Model~3. During most of the evolution, the flow in
the primary bar, outside the interface, is steady and the shock strength and
shape are nearly independent of time. This quasi-steady state exists after
$t\sim 10$ and till the end of the numerical simulations at $t=30$. Our
subsequent analysis is limited to this time only.    

In the second region, the bar-bar interface, the flow has a time-dependent
character due to the perturbative effects of the secondary bar and changing
background potential, and naturally correlates with the position angle of the
secondary bar with respect to the primary bar, especially in Model~2
due to a more pronounced small-scale bar (Fig.~2). To get some insight into
the flow across this zone, and specifically across the corotation of the
secondary bar, we have subdivided the azimuthal dependence of the gas flow
into $\pm 45^\circ$ with the major axis of the primary bar and $\pm 45^\circ$
with its minor axis (Fig.~3a). We have Fourier analyzed the time dependence of
the mass inflow rates and found the trace of the beat frequency of the
secondary bar in Model~1, which in the frame of the primary bar should appear
at $f \sim 2.3$ and shows up at 2.23 (Fig.~3b), althought other frequencies
have substantial power as well. Model~2, having a stronger bar influence at
the interface, exhibits a higher Fourier amplitude at exactly the beat frequency
$f=2.3$, clearly identified with the secondary bar tumbling and
with less power from other frequences. 

So the flow across the bar-bar interface depends upon the strength of the
secondary bar. It ranges from more chaotic, for a relatively weak perturbation of
the secondary bar in Model~1, to a more regular one in Model~2.
The corresponding mass influx rate is of the order of $0.3\!\ M_{gas,9}\ {\rm
M_\odot\ yr^{-1}}$, where $M_{gas,9}$ is the total gas mass in the disk in
units of $10^9$~M$_\odot$. On the average, the inflow proceeds through the
broad region along the primary bar minor axis, while an outflow (albeit at a
smaller rate) is directed along its major axis. The reason for this behavior is
that the inflow is driven mainly along the large-scale shocks penetrating the
bar-bar interface from the primary bar. At the same time the outflow is
detected at angles which do not encompass the large-scale shocks. The net
effect is clearly an inflow across the corotation of the secondary bar, as
indicated in Fig.~3a.

As a caveat, gas, which is repelled by the secondary bar along the major axis
of the primary bar, is found to enter large-scale shocks while still moving
out. This should aggravate, at least in principle, the mixing of material with
a different angular momentum. Such mixing will lead to an increased
inflow along the shocks and even less rotational support upon the entrance to
the secondary bar. No attempt was made to quantify this effect.

The gas response at the interface and at smaller radii differs substantially
among all three models. As expected, the single bar Model~3 shows no signs
of further gas evolution which stagnates in the vicinity of the ILR in a 
nuclear ring. Fig~2 displays a single frame of gas distribution for this model 
at $t=15$, which
is also characteristic of later times. Hence we have refrained from showing
other frames. This is not the case for models with secondary 
bars which drive the gas inward, towards  
smaller radii as can be seen from Fig.~2.  The gas from
the primary bar is crossing the bar-bar interface
along the large-scale shocks, settling well inside
the nuclear bar whose length is indicated by the straight line. 
Specifically, it falls towards the third region, within $r <
0.02$ (Model~1) and $<0.04$ (Model~2). This region 
shows a very relaxed flow at all times, with uniform dissipation (well
below the maximum dissipation in the large-scale shocks), and no evidence for
grand-design shocks.

An important difference between Models~1 and 2 is the absence of a nuclear
ring in the latter model. While Model~1 shows a well developed ring at all
times, made out of two tightly wound spirals (Fig.~4a, right column). Model~2 
develops a pair of open
spiral shocks, but only when the bars are closely aligned (Fig.~4b, right column).
 
In order to understand the pattern of shock dissipation in nested bars, we
show the time evolution of the gas which has more than the average dissipation 
rate within the central kpc ($r=0.1$) at different times (Figs.~4). This
scale allows one to separate the incoming large-scale shocks from those driven
by the secondary bar. 

We first note that two systems of spiral shocks occur in Models~1 and 2,
each associated with their corresponding bar. The large-scale (hereafter
`primary') shocks, which have been discussed above, normally extend to the
minor axis of the primary bar in both Models~1 and 2, e.g., Fig.~4a at
$t=28.2$, Fig.~4b at $t=12.2$, or Fig.~4c at $t=28.1$. At these times the
secondary bar is nearly orthogonal to the primary one. As it continues to
tumble, the primary shocks extend deeper into the small bar. Sometime
before both bars are perpendicular, the outer shocks detach from the small bar, which
is left with additional pair of trailing shocks. This effect is especially
pronounced in Fig.~4a between $t=28.1$ and 28.2, in Fig.~4b between the
times $t=12.1$ and 12.2, and in Fig~4c between $=28.0$ and 28.1.

\figurenum{4b}
\begin{figure}[ht!!!!!!!!!!!!!!!!!!!!!!!!!]
\vbox to5.75in{\rule{0pt}{5.75in}}
\includegraphics{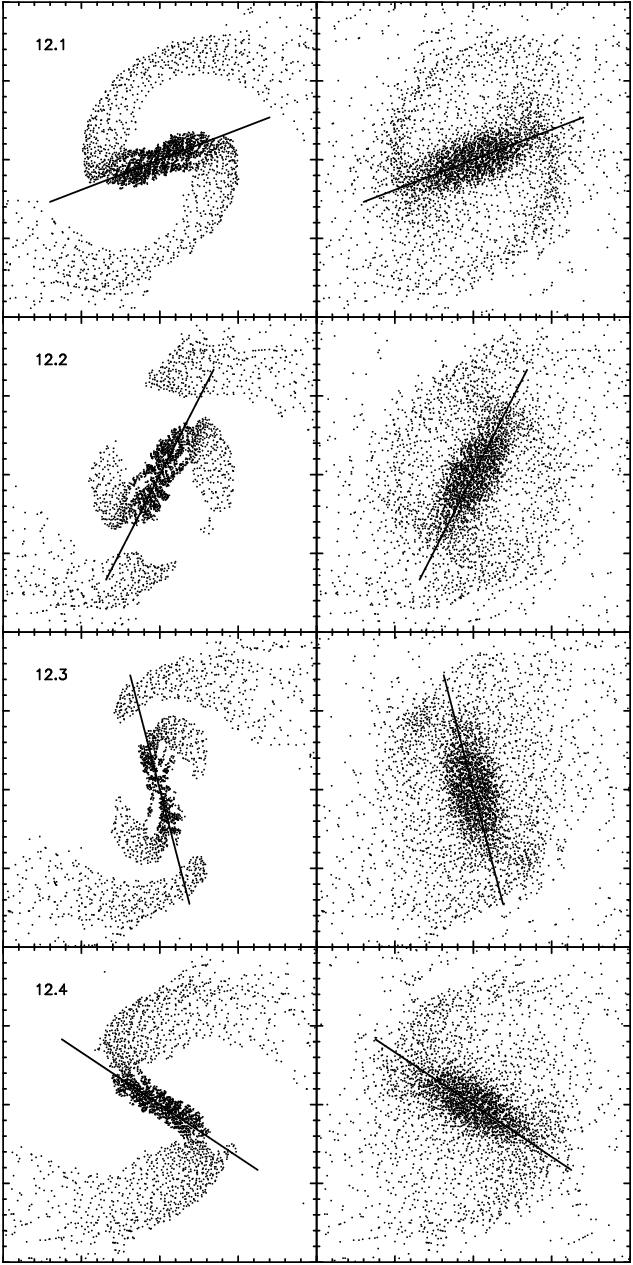}
\caption{Pattern of shock dissipation (left) and density evolution (right) in
the central kpc ($r=0.1$) of Model~2 at early times, shown in the frame of reference
of the primary bar (horizontal). Position of the secondary bar  and its
length are indicated with a straight line. All rotation is counter-clockwise.
The particles on the left are those having greater than average dissipation
rate which is given by the time derivative of the nonadiabatic component of
internal energy. Note the broadly shocked region of the bar in the innermost
0.05. Also visible are two dissipative systems associated with the large-scale
shocks in the primary bar and with the trailing shocks in the secondary bar.}
\end{figure}
\figurenum{4c}
\begin{figure}[ht!!!!!!!!!!!!!!!!!!!!!!!!!]
\vbox to6.0in{\rule{0pt}{6.0in}}
\includegraphics{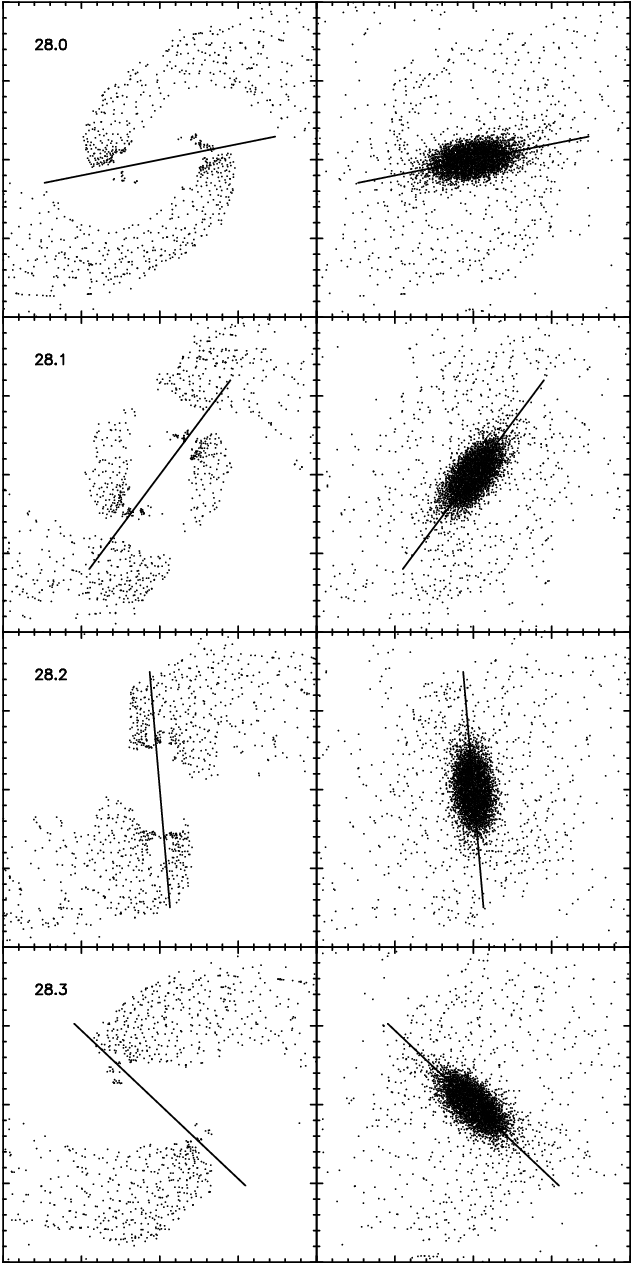}
\caption{Pattern of shock dissipation (left) and density evolution (right) in
the central kpc ($r=0.1$) of Model~2, shown at late times. See Fig.~4b
caption for further details.}
\end{figure}                

To summarize, the interaction between these shock systems shows attachment when
the bars are aligned with each other and detachment when they are 
perpendicular. The shapes of the secondary shocks depend on the angle between the bars.

The most dramatic difference between the models comes from the gas evolution
in the third region, deep inside the secondary bar. In Model~1, the gas
settles within the central $r\approx 0.02$ (i.e., 200\,pc) where it experiences 
very little dissipation, compared to the outer shocks. In fact, we observe a
kind of ``limb brightening'' at the edge of this bar. This is seen in
Fig.~4a  as an enhanced density of above-the-average dissipating particles
outside the ovally-shaped central region. The reason for this is that the gas
joins the bar from all azimuths. In Model~2, the interior of the gaseous
bar is uniformly dissipative at earlier times (Fig.~4b). The source of this
dissipation is the small-scale shocks which are typically perpendicular to the
major axis of the secondary bar. This is not the
type of  `centered' shocks observed by Athanassoula (1992) as it uniformly
encompasses the inner bar and the width of this dissipation zone is roughly
equal to the minor axis of the bar. Note, that the dark shade here does {\it not}
mean increased dissipation rate per particle, it only reflects the high density 
of the SPH particles. 
At later times, this dissipation rate
decreases sharply (Fig.~4c). 

We have looked more carefully into the central dissipation of Models~1
and 2 by using logarithmically grey-scaled maps (Figs.~5a,b). Both the ``limb
brightening'' and the broad dissipation can be observed in detail. While some
some spatial dependency of the dissipation morphology can be seen in Fig.~5b 
of Model~2, no large-scale shocks are present.

\figurenum{5}
\begin{figure}[ht!!!!!!!!!!!!!!!!!!!!!!!!!]
\vbox to4.5in{\rule{0pt}{4.5in}}
\includegraphics{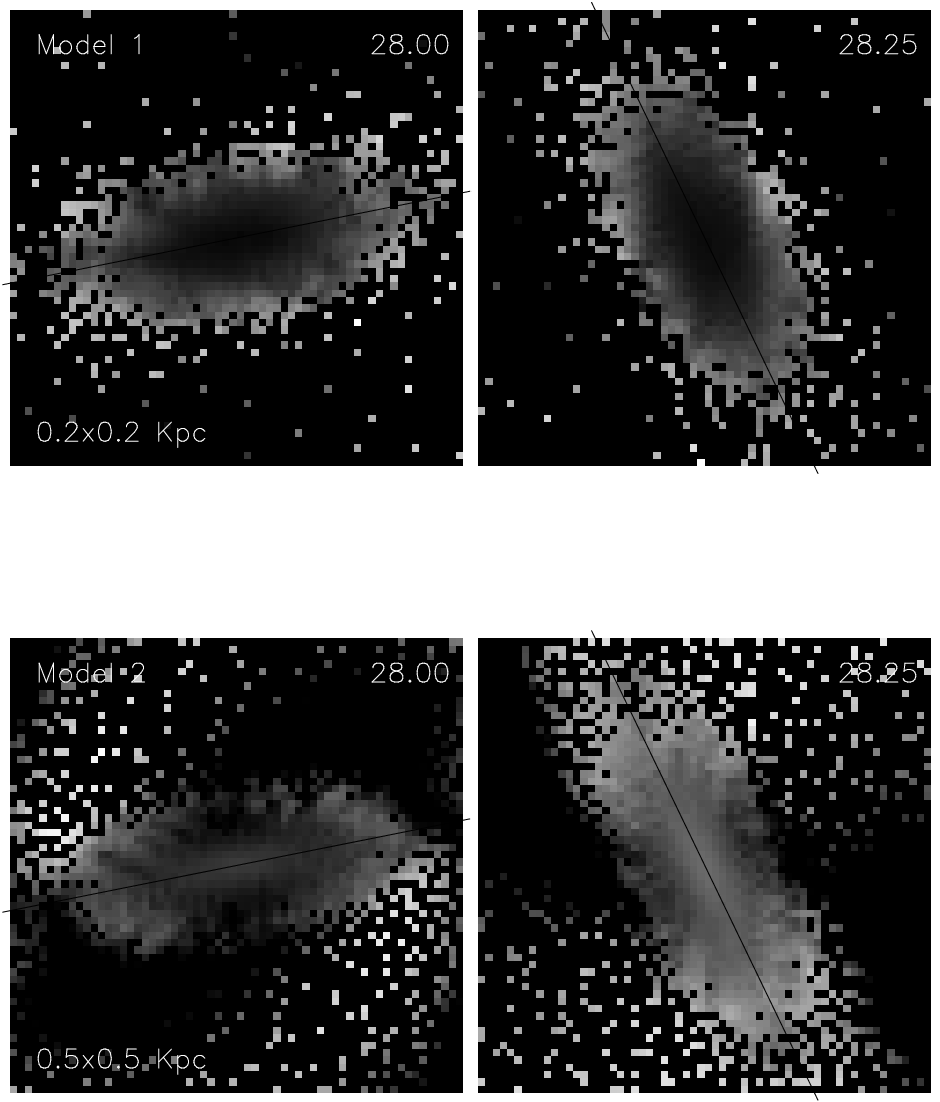}
\caption{Logarithmic grey-scale map of shock dissipation in the interior of the
secondary bar aligned at two different angles with respect to the primary bar
(horizontal): ($a$) Model~1 (the inner 0.02), ($b$) Model~2 (the inner 0.05).
The grey level is the logarithm of the viscous dissipation rate averaged over
a pixel and scaled between $0.01-1.0$ of the maximum rate, i.e., darker colors
represent less dissipation. It is given by the time derivative of the
nonadiabatic component of internal energy. The orientation of the secondary
bar is given by the solid line. All rotation is counter-clockwise.}
\end{figure}  
 
\section{Discussion: Gas Dynamics in Nested vs Single Bars}
  
This paper deals with the gas flow in dynamically-decoupled nested bars, 
focusing on two central issues, namely ($i$) is the gas capable of crossing the
bar-bar interface in the nested bar systems, and ($ii$) do offset dust lanes
form in the secondary bars. We first comment on the principal differences
between the flow in the {\it  single} large-scale bars and in the {\it nested}
bars. 

\subsection{Gas flow across the bar-bar interface}

In the presence of only one bar with a double ILR, the gas accumulates
between the resonances in the form of two nuclear rings, as a result of shock
focusing. These rings may or may not merge due to hydrodynamical interaction,
depending on the width of the resonance region, shape of the underlying
potential and the level of star formation, unrelated to the importance of
gas self-gravity. The slowdown of gas evolution at the resonances has been
known for some time. Past numerical simulations (e.g., Combes \& Gerin 1985;
Piner, Stone \& Teuben 1995), however, did not catch the possibility of a
non-self-gravitational dynamical runaway which can develop in the nuclear
rings (Heller, Shlosman \& Englmaier 2001). With self-gravity, the gas
accumulation near the ILRs is subject to global instabilities, and
further inflow is expected to be accompanied with star formation
(Knapen et al. 1995b; Shlosman 2001). 

As argued by Pfenniger \& Norman (1990), the gas flow in nested bars has
difficulty in crossing the bar interface, being repelled there unless
self-gravitational effects in the gas develop. One can understand this by
noting that the effective potential shape corresponds to a rim at the
bar corotation from which the gas is forced to move away, both inwards or outwards.
Our modeling shows that the gas flow, in fact, has no difficulty in crossing 
this resonance, and that this crossing is facilitated by the large-scale shocks
which penetrate the region (roughly) along the minor axis of the primary bar,
as shown in Figs.~4. We do observe a relatively insignificant outflow along the
major axis of the primary bar, just outside the corotation of the small bar.
But this outflow is completely offset by the inflow along the large-scale
shocks of the main bar, coming in along the minor axis of the bar. This
phenomenon was not captured by the ``axisymmetric'' analysis of Pfenniger \& 
Norman. 

The flow in single and double bars is found to be remarkably similar {\it outside}
the bar-bar interface zone in all three models. This similarity, however, 
ends at the interface. Model~3 forms a nuclear ring made up of tightly wound
spiral shocks and very little action takes place inside this ring. Model~1,
with the short secondary bar, forms a similar ring which apparently is able to
survive the perturbing action of the small bar. Note that the ring radius is
about 0.04--0.05, and sits just at the bar edge, well inside its corotation.  
In contrast, the nuclear ring is being constantly disrupted and is not
present most of the time in Model~2, in which the small bar extends close to its
own corotation. Except when the bars are nearly aligned, the inner bar is
situated in a disk-like envelope of gas, e.g., at times $t=12.2$ and 12.3 in 
Fig.~4b. The incoming large-scale shocks are
most prominent when the bars are aligned and disappear at other angles. 

An important observational corollary is that nuclear rings are expected to be
absent in decoupled nested bars, unless the secondary bars are very
centrally concentrated and/or short compared to their corotation.

The shape of the gas response inside the small bar also differs between the
models. Model~1 exhibits a small gaseous bar of about 1/3 of the imposed
stellar potential, and the shape is elliptical at all times (Fig.~4a). The 
gas is unable to settle down in the outer part of the bar due to the time
dependent gravitational potential there. In addition, the characteristic
diagram for this bar (Fig.~1b) has unstable and intersecting $x_1$ orbits
for $r > 0.03$, and so the subsequent gas behavior comes as no surprise.

In Model~2, the evolution is more complicated as the $x_1$ orbits are stable
but the variability of the gravitational potential is more severe. In the early
stage of the gas inflow, it is filling up to 90\% of the secondary bar, and
the shape of the gaseous bar changes from being oval to rectangular (when it
leads the primary bar by about $45^\circ$) (Fig.~4b). This shape corresponds 
to the shape of the outer orbits in the small bar, close to its Ultra-harmonic
resonance. At later times, the gaseous bar is about half of the size of the
secondary bar background potential and its shape is oval as in the Model~1.

\subsection{Absence of offset grand-design shocks in secondary bars}

Numerical simulations presented here demonstrate that the gas dynamics of the
decoupled secondary bars differs substantially from that of the primary or
single bars. A number of factors contribute to this, the first of which is the
time-dependent nature of the gravitational potential in the nested bar system 
due to the  distinct rates of bar tumbling during the dynamically-decoupled
phase. The second factor is the gas injection into the secondary bar which
proceeds through the primary large-scale shocks penetrating the bar-bar
interface. No such phenomenon is operating across the corotation of the
primary bars, which is rather depopulated of gas. The above factors are
accompanied by a large amount of dissipation and the subsequent inability of
the gas to settle in the outer half of the secondary bar. This raises the
interesting question of whether secondary bars extend to their corotation, as
their large-scale counterparts are believed to do. Based on the numerical
simulations, we infer that secondary bars are more centrally concentrated
than primary or single bars, and that the gas distribution in these bars does
not extend to their corotation radii. Even if the $x_1$ orbits can, in 
principle, be
found at energies close to the Ultra-harmonic resonance, in many cases they
are unstable, self-intersect and are unable to trap regular orbits around
them. These orbits cannot support the gaseous component as well. 

The third factor which differentiates the secondary from other type
of bars is their fast pattern speed which does not allow for secondary
ILRs to form inside the bar. If the decoupled phase of nested bars is 
short-lived, the quadrupole interaction between the bars will not be able to
brake the small bar and form the ILRs. However even in the case of a
long-lived decoupled phase we do not expect the nuclear bars to slow down. 
In fact, the gas inflow across the interface and the resulting central concentration 
speed up the bar, as can be inferred from numerical experiments with
Sticky Particles (Shlosman \& Noguchi 1993) and SPH (Heller \& Shlosman
1994) hydrocodes. Because of the fast rotation, the $x_1$ orbits deep
inside the secondary bars are round, with no end-loops or needle shapes.
As noted by Athanassoula (1992), for the shocks to exist, the curvature of the
$x_1$ orbits at apocenters must be sufficiently large, or they must have end-loops.
The low Mach number gas flow is well organized here and capable of following
these orbits with little dissipation, as shown in Figs.~4b,c.

Indeed, the nonlinear orbit analysis (Fig.~1) shows that the main orbits 
aligned with the secondary bar, $x_1$ have a mild ellipticity and no end-loops.
This result is rather robust and holds despite the extreme axial
ratio, $4:1$, used here. No offset large-scale shocks form under these
conditions and hence no offset dust lanes are expected either.  

Hence, it is highly probable that decoupled secondary bars avoid their ILRs 
because of their high pattern speeds. When the ILRs are absent in a
large-scale bar, the offset shocks weaken and recede to the major axis of the
bar becoming ``centered'' (e.g., Athanassoula 1992). They do not disappear
completely because the underlying stellar periodic orbits have either
end-loops, are pointed, or have large curvature at the ends, forcing the gas
to shock there. These orbital shapes result from the slower rotation of large
bars compared to the nuclear ones. Despite the fact that such weak centered
shocks can exist theoretically, only two examples have been found out of more 
than a hundred barred galaxies analyzed by  Athanassoula (1992),
one of which is the dubious case of NGC~7479. During the last decade
only one more potential example has been added to this list (Athanassoula,
private communication). We conclude that centered shocks are very rarely
observed, possibly because of being so weak. 

No `classical' centered shocks have been observed in our numerical simulations.
Instead we find that the inner half of the secondary bars show a rather 
uniform dissipation during the early stages of the gas inflow (Fig.~4b), which
then sharply decreases at latter times (Figs.~4c, 5a,b). This dissipation is
always small compared to dissipation in the primary large-scale shocks.
It is interesting that a nuclear Lindblad resonance introduced by a central
supermassive black hole should in principle lead to trailing spiral shocks,
but at very small radii, $\sim 10-50$~pc (Shlosman 1999; 2001). 

It is important that both short and long (with respect to their corotation)
nuclear bars show basically the same distribution of the gas component. The
bar-bar interface (i.e., nuclear bar corotation) is almost always
depopulated of gas because of the absence of non-intersecting
orbits there. This conclusion holds for gas-dominated and star-dominated 
nuclear bars and has interesting implications
which will be discussed elsewhere.

Knapen et al. (1995b) have analyzed the shock dissipation in a self-consistent
gravitational potential of `live' stars and gas {\it before} the decoupling
phase, when both bars tumble with the same pattern speeds, and when the gas
self-gravity is accounted for. No offset shocks have been found in this
configuration either. Dust lanes, therefore, cannot be used to search 
for nuclear bars, as proposed by Regan \& Mulchaey (1999) and Martini \& Pogge 
(1999). Furthermore, the alternative method of NIR isophote fitting
--- a reliable approach in detecting large-scale bars, has its own difficulties
when applied to nuclear bars. This has been shown by Laine et al. (2001) for the
largest to-date matched sample of 112 Seyferts and non-Seyferts. The main difficulty in
fitting the isophotes comes from localized and distributed sites of star 
formation, especially pronounced within the central kpc. This results in
a substantial underestimate of the nuclear bar fraction and cannot be used
reliably in order to analyze e.g., the role of nuclear bars in fueling of 
active nuclei. However, recent work 
by Martini et al. (2001), without invoking a matched control sample, is 
doing just this.  
Clearly, the most promissing method in detecting the nuclear bars is
2D spectroscopy of the central kpc
revealing the underlying kinematics.  

An important issue is the fate of the gas accumulating in the inner parts of 
the secondary bars. Under the observed conditions in our numerical
simulations (gas masses and surface densities) the gas self-gravity should have
a dominating effect on its evolution.
Shlosman (2001) has described the work of Englmaier \& Shlosman (1999, 
unpublished)
who studied the dynamical stability of nuclear rings
under similar conditions and found that global self-gravitational modes with
$m=2$ and 4 are rapidly amplified into the non-linear regime. The gas loses
its rotational support and falls towards the center, feeding the central
supermassive black hole at peak rates and increasing their mass tenfold.
Of course a big unknown is the concurrent star formation, which was neglected
in these simulations. We note, however, that the star formation is unknown
anywhere to reach so high an efficiency, that it would be able to  halt this
runaway collapse to the center. Clearly, conditions obtained at
the end of our present numerical simulations of gas flow in nested bars are
extreme and will be destabilized by the self-gravitational
instabilities which will extract angular momentum from the gas on a
dynamical timescale. It is important, therefore, that a `live' gravitational
potential of the bars immersed in the disk be used in order to get a qualitative and
quantitative picture of further evolution in the central kpc of disk galaxies.

We conclude that no large-scale shocks and consequently no offset dust lanes
will form inside secondary nuclear bars either when they are dynamically coupled 
and spin with the same pattern speeds as the primary bars, or dynamically decoupled,
spinning much faster. Two main factors, the time-dependent gravitational
potential and the nature of the gas flow deep inside the bar, prevent the
formation of these dust lanes.  The time-dependent, ``anisotropic''  gas
inflow across the ILR/corotation interface found here is a completely new
phenomenon inherent to nested bars, because in the single barred systems, no
such inflow is possible at all {\it  when the gas accumulation is small and
its self-gravity is negligible}. The fate of the gas settling inside the
nuclear bars cannot be decided without invoking global self-gravitational
effects in the gas which will completely change the nature of the flow there.
Star formation most probably will play an important role here, but because of
its expected low efficiency,  it is doubtful it would be capable of halting
the inflow.

\acknowledgments

We thank Lia Athanassoula and Peter Englmaier for numerous discussions and the
organizers of INAOE workshop on {\it Disk Galaxies: Kinematics, Dynamics and
Perturbations} for supporting a prolonged visit during which this work was
concluded. Supported in part by NASA grants NAG 5-10823, NAG 5-3841,
WKU-522762-98-6 and HST GO-08123.01-97A to I.S.

\end{document}